\documentclass[conference]{IEEEtran}
\IEEEoverridecommandlockouts
\usepackage{cite}
\usepackage{amsmath,amssymb,amsfonts}
\usepackage{algorithmic}
\usepackage{graphicx}
\usepackage{textcomp}
\usepackage{xcolor}
\def\BibTeX{{\rm B\kern-.05em{\sc i\kern-.025em b}\kern-.08em
    T\kern-.1667em\lower.7ex\hbox{E}\kern-.125emX}}
\begin{document}

\title{Machine Learning-Based GPS Multipath Detection Method Using Dual Antennas

\thanks{This research was supported by the Unmanned Vehicles Core Technology Research and Development Program through the National Research Foundation of Korea (NRF) and the Unmanned Vehicle Advanced Research Center (UVARC) funded by the Ministry of Science and ICT, Republic of Korea (2020M3C1C1A01086407).
This work was also supported by Institute of Information \& Communications Technology Planning \& Evaluation (IITP) grant funded by the Korea government (KNPA) (2019-0-01291).}
}

\author{\IEEEauthorblockN{Sanghyun Kim}
\IEEEauthorblockA{\textit{School of Integrated Technology} \\
\textit{Yonsei University}\\
Incheon, Korea \\
sanghyun.kim@yonsei.ac.kr}
\and
\IEEEauthorblockN{Jungyun Byun}
\IEEEauthorblockA{\textit{School of Integrated Technology} \\
\textit{Yonsei University}\\
Incheon, Korea \\
jungyun.byun@yonsei.ac.kr}
\and
\IEEEauthorblockN{Kwansik Park} 
\IEEEauthorblockA{\textit{Korea Aerospace Research Institute} \\
Daejeon, Korea \\
kspark6469@kari.re.kr}
}

\maketitle

\begin{abstract}
In urban areas, global navigation satellite system (GNSS) signals are often reflected or blocked by buildings, thus resulting in large positioning errors.
In this study, we proposed a machine learning approach for global positioning system (GPS) multipath detection that uses dual antennas.
A machine learning model that could classify GPS signal reception conditions was trained with several GPS measurements selected as suggested features. We applied five features for machine learning, including a feature obtained from the dual antennas, and evaluated the classification performance of the model, after applying four machine learning algorithms: gradient boosting decision tree (GBDT), random forest, decision tree, and K-nearest neighbor (KNN).
It was found that a classification accuracy of 82\%–96\% was achieved when the test data set  was collected at the same locations as those of the training data set.
However, when the test data set was collected at locations different from those of the training data, a classification accuracy of 44\%–77\% was obtained.
\end{abstract}

\begin{IEEEkeywords}
multipath detection, global positioning system, machine learning, dual antennas
\end{IEEEkeywords}

\section{Introduction}
Global navigation satellite systems (GNSSs), particularly the global positioning system (GPS) of the United States of America, play an essential role in various aspects of our lives \cite{Park2020800, Yoon20, Jeong2020958, Shamaei21, Kim2020796, Maaref20, Park2020824, Son20191828, Han2019, Rhee21:Enhanced, Schmidt20, Park2021919, Braasch19, Park2018387, Kim2019, Knoop17, Kang21:Indoor, Sun21:Markov, Lee22:Optimal}. Recently, as autonomous vehicles, such as unmanned aerial vehicles (UAVs), have emerged, the use of GNSS has increased further \cite{Kim2020784, Sun2020889, Moon2019157, Causa21, Moon21, Lee20191187, Moon22:Fast, Lee2018:Simulation, Moon202013, Savas21, Moon2019258, Moon20181530, Park21:Indoor, Moon22:Speeding, Moon22:Sample}.

However, in urban areas, it is difficult to achieve high positioning accuracy with the GNSS because the signals get reflected by buildings in many cases \cite{Lee2020:Preliminary, Lee2020939, Jia21:Ground, Lee20202347, Kim22:First, Lee22:SFOL, Jeong21:Development, Kim21:GPS}. Using the reflected GNSS signals for positioning could result in a positioning error greater than 100 m \cite{MacGougan02}. From the perspective of signal reception, the GNSS signal can be received in the following three cases: when only line-of-sight (LOS) signal is received (that is, \textit{LOS-only condition}), only non-line-of-sight (NLOS) signal is received (that is, \textit{NLOS-only condition}), and both LOS and NLOS signals are simultaneously received (that is, \textit{LOS+NLOS condition}). 
Among these conditions, the performance degrades only in the NLOS-only, and LOS+NLOS conditions. Therefore, various methods for detecting multipath signals (that is, NLOS-only or LOS+NLOS signals) and mitigating the corresponding GNSS errors are being actively studied \cite{Massarweh20, Shen20, Kubo20}.

The traditional method of detecting multipath signals uses the carrier-to-noise-density ratio ($C/N_0$) measurement of the GNSS signal \cite{Groves10}. By defining a threshold for the $C/N_0$ values, a signal with $C/N_0$ greater than the threshold is classified as an LOS-only signal, and a signal with $C/N_0$ lesser than the threshold is classified as an NLOS-only or LOS+NLOS signal. 
Similarly, there are numerous software-based detection algorithms for GNSS measurements, such as the Doppler shift and code-minus-carrier values \cite{Xu19, Lee20}. 
Additionally, there are hardware-based detection methods that use a special antenna \cite{Closas11}, a sky-pointing camera  \cite{Bai20}, and a method that uses a three-dimensional (3D) model of a city and ray-tracing techniques \cite{Kumar14}. Furthermore, GNSS multipath signal detection techniques using machine learning approaches have been emerging recently \cite{Sun21, Suzuki20, Munin20}.

Various machine learning techniques have been used for classifying GNSS signal reception conditions, and their multipath detection performances were better than those of traditional methods \cite{Guermah18, Sun19, Hsu17}. 
The performance of a machine learning technique is greatly influenced by the features that are used, and the algorithms that are applied to it. Guermah \textit{et al.} \cite{Guermah18} proposed a method that used the $C/N_0$ difference value obtained by a dual-polarized antenna, as a feature for the machine learning model. Sun \textit{et al.} \cite{Sun19} classified LOS-only, NLOS-only, and LOS+NLOS signals by applying the gradient boosting decision tree (GBDT) algorithm, and used $C/N_0$ values, pseudorange residuals, and satellite elevation angles as features for machine learning. 

In this study, we proposed a machine learning-based GPS multipath detection method that uses dual antennas.
Using dual antennas instead of a single antenna helps to obtain more diverse GPS measurements, and utilize them as features for machine learning.
Resultantly, it is expected that the performance of classifying signal reception conditions would be improved.
Using more than two antennas might be more effective, but it has disadvantage in terms of the size and cost.

We used four features (namely, $C/N_0$, temporal difference of $C/N_0$, difference between delta pseudorange and pseudorange rate, and satellite elevation angle) that could be obtained from a single antenna, and one feature (namely, double-difference pseudorange residual) that could be obtained from dual antennas.
Four algorithms were applied as the machine learning algorithms in this study, and their performances were compared.
There has been a study on GNSS multipath detection using dual antennas \cite{Chen21}. However, unlike our study, it was based on a statistical approach, and not a machine learning approach.
Compared to statistical methods, machine learning methods can identify dependencies in data sets that cannot be modeled mathematically \cite{Siemuri21}.
We classified signal reception conditions by a machine learning method that utilizes five features, whereas the existing method \cite{Chen21} classified the conditions by a statistical method that uses a single metric.

\section{GPS Signal Collection and Labeling}
\subsection{GPS Signal Collection}
\begin{figure}
  \centering
  \includegraphics[width=0.8\linewidth]{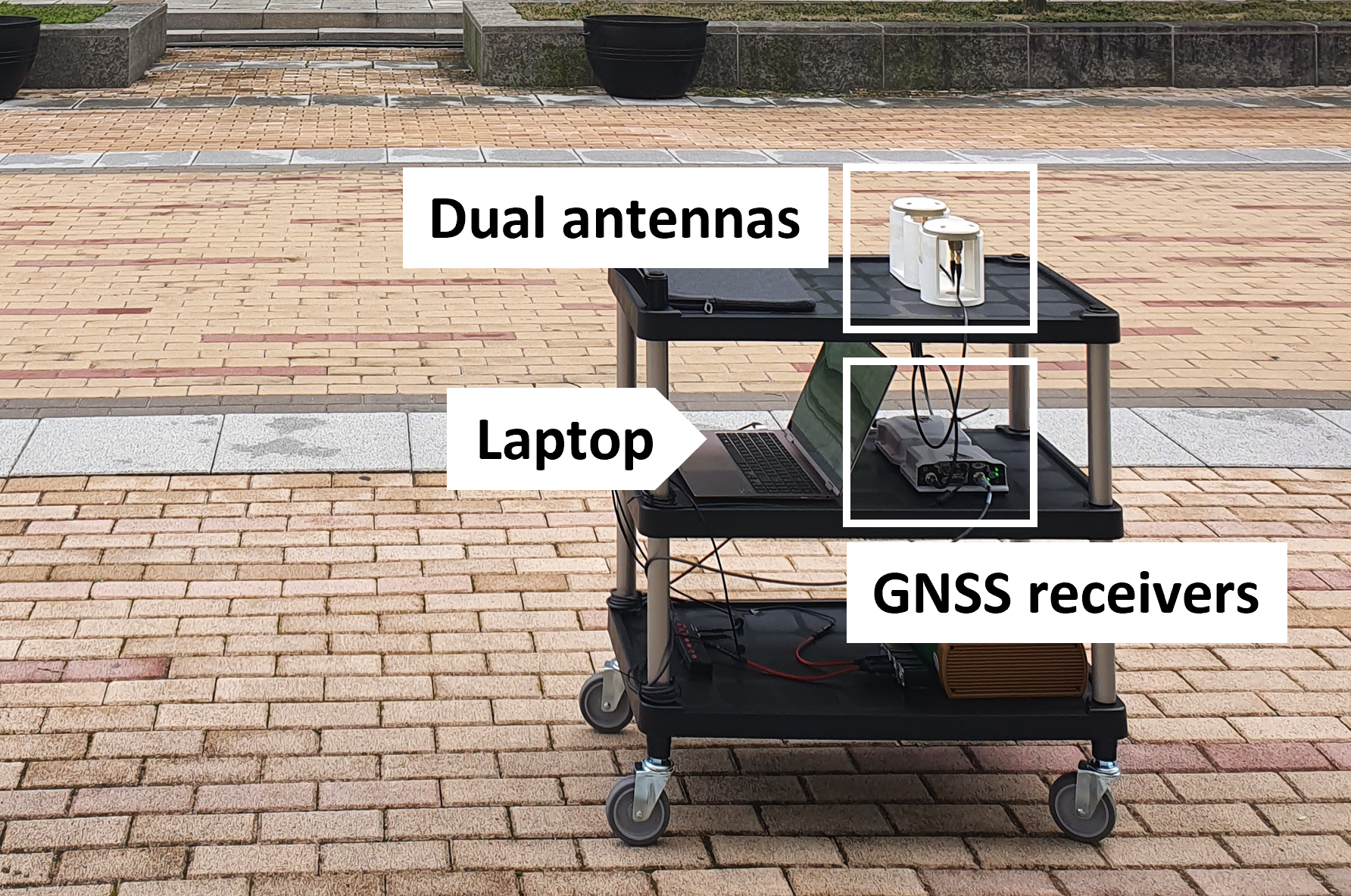}
  \caption{Hardware setup for GPS signal collection}
  \label{fig:Hardware}
\end{figure}

 Fig.~\ref{fig:Hardware} shows the GPS signal acquisition hardware that was set up to collect the large amount of GPS data for machine learning. It consisted of two right-hand circularly polarized (RHCP) antennas produced by Antcom, two NovAtel PwrPak7 GNSS receivers, and a laptop for storing raw GPS data. The distance between the two antennas was maintained at 19 cm, which was equal to the wavelength of the GPS L1 signal. We collected GPS L1, L2, and L5 signals at intervals of 1 s, in a static environment.

\begin{figure}
  \centering
  \includegraphics[width=1.0\linewidth]{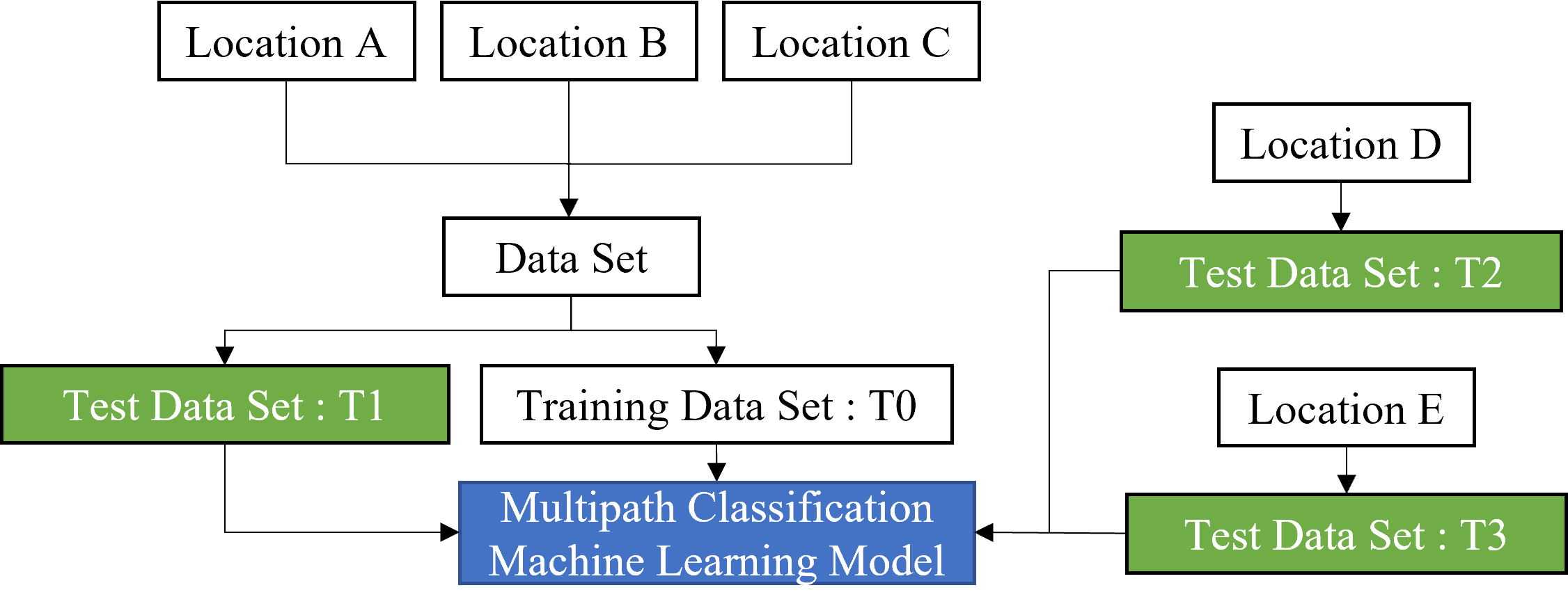}
  \caption{Training and test data set formulation based on the collected data at five locations}
  \label{fig:Dataset}
\end{figure}

Fig.~\ref{fig:Dataset} illustrates the training and test data sets. The GPS signals were collected at five locations within Yonsei University, Incheon, Korea, which were labeled as A, B, C, D, and E, all having similar multipath environments.
The data collected at three of these locations (that is, A, B, and C) were combined into one data set so that the number of LOS-only, NLOS-only, and LOS+NLOS samples within the data set could be similar to each other.
Then, the same number of LOS-only, NLOS-only, and LOS+NLOS samples were randomly selected from the combined data set to form the training data set, T0, and the test data set, T1.
Table \ref{table:LabelResult} lists the number of data samples included in each data set.

The data collected at the remaining two locations (that is, D and E) were used only for testing. T2 and T3 in Table \ref{table:LabelResult} represent the data set collected at locations D and E, respectively.

\begin{table}
\centering
\caption{The number of data samples for each data set}
\renewcommand{\arraystretch}{1.5}
\begin{tabular}{l||cccc}\hline
 GPS data set & T0 & T1 & T2 & T3\\
 \hline
 Total samples & 7500 & 7500 & 6311 & 11575 \\
 NLOS-only samples & 2500 & 2500 & 2038 & 3608 \\
 LOS-only samples & 2500 & 2500 & 2195 & 4033 \\
 LOS+NLOS samples & 2500 & 2500 & 2078 & 3934 \\
 \hline
\end{tabular}
\label{table:LabelResult}
\end{table}

\subsection{Data Labeling Based on Ray-Tracing Simulation}

It is essential to label the true signal reception conditions on the GPS data collected for machine learning. In this study, labeling was performed through ray-tracing simulation, using a commercial ray-tracing software called Wireless InSite \cite{WirelessInsite}. 
The ray-tracing simulator displays the propagation paths of the signal from the GPS satellite to the receiver, based on the 3D building information around the receiver.
We used a commercial 3D building map provided by 3dbuilings \cite{3Dbuildings}, and calculated the locations of the GPS satellites based on almanac data.


\section{Feature Selection and Machine Learning Algorithms}
\subsection{Features to Classify the Signal Reception Conditions}

The features used in this study were as follows: $C/N_0$, temporal difference of $C/N_0$, difference between delta pseudorange and pseudorange rate, satellite elevation angle, and double-difference pseudorange residual.

\begin{itemize}
\item $C/N_0$: $C/N_0$ is defined as the carrier-to-noise density ratio. It is a parameter containing received signal strength information, and its unit is dB-Hz. Generally, since the propagation loss increases when a signal is reflected by an obstacle, the $C/N_0$ value of the LOS-only signal is larger than that of the NLOS-only signal.

\item Temporal difference of $C/N_0$: This parameter is denoted as $\Delta C/N_0$, which indicates the change in $C/N_0$ between two consecutive epochs. In a static environment, $C/N_0$ values of NLOS-only or LOS+NLOS signals can increase with respect to time, based on the characteristics of the receiver tracking loop \cite{Hsu17}.

\item Difference between delta pseudorange and pseudorange rate: This parameter represents the consistency between the pseudorange measurement and the Doppler shift. It has been proposed as a feature in \cite{Hsu17}, and can be calculated as follows:
\begin{equation}
  \Delta \rho - \dot \rho \cdot \Delta t
  \label{eqn:PRRate}
\end{equation}
where $\Delta \rho$ is the change in pseudorange between two epochs; $\dot \rho$ is the pseudorange rate, which is related to the Doppler shift; and $\Delta t$ is the time difference between the two epochs.

\item Satellite elevation angle: Generally, the higher the elevation angle of the satellite, the lower is the probability that the signal is reflected or blocked by a building.

\item Double difference pseudorange residual: The above four features can be obtained from a single antenna, and they have been proposed in previous literature. However, dual antennas are necessary to obtain the double difference pseudorange residual, which was proposed as a feature for multipath detection for the first time in this study. 
This feature can be calculated as follows: 
\begin{equation}
\begin{split}
  \rho_{12,\mathrm{residual}}^{kj}
  & = \rho_{12,\mathrm{m}}^{kj} - \rho_{12,\mathrm{e}}^{kj} \\
  & = \{(\rho_{1,\mathrm{m}}^{k}-\rho_{2,\mathrm{m}}^{k}) - (\rho_{1,\mathrm{m}}^{j}-\rho_{2,\mathrm{m}}^{j})\} \\
  & \, \; \quad - \{(\rho_{1,\mathrm{e}}^{k}-\rho_{2,\mathrm{e}}^{k}) - (\rho_{1,\mathrm{e}}^{j}-\rho_{2,\mathrm{e}}^{j})\}
  \label{eqn:DDPR}
\end{split}
\end{equation}
where $\rho_{12,\mathrm{m}}^{kj}$ and $\rho_{12,\mathrm{e}}^{kj}$ are the measured and expected double difference pseudorange values between satellites $k$ and $j$, and receivers 1 and 2, respectively. The expected pseudorange is the distance between the satellite and the receiver, where the satellite and receiver positions are obtained based on the ephemeris data and the receiver position fix, respectively.
Since this feature was calculated based on the signals from two satellites, the satellite with the highest elevation angle was selected as the reference satellite for the given epoch and the double difference pseudorange residual was calculated between each satellite and the reference satellite.

If the received signal was NLOS-only or LOS+NLOS, its propagation path from the satellite to each of the two antennas would be noticeably different, which caused the absolute value of the double difference pseudorange residual to increase. Therefore, this feature can be used for the classification of signal reception conditions. 
\end{itemize}

\subsection{Machine Learning Algorithms}
The four machine learning algorithms used in this study were GBDT, random forest, decision tree, and K-nearest neighbor (KNN). We used the algorithms in a machine learning library called scikit-learn \cite{Pedregosa11}, and performed hyperparameter tuning using the GridSearchCV function included in scikit-learn.

\section{Results}
\begin{figure}
  \centering
  \includegraphics[width=0.8\linewidth]{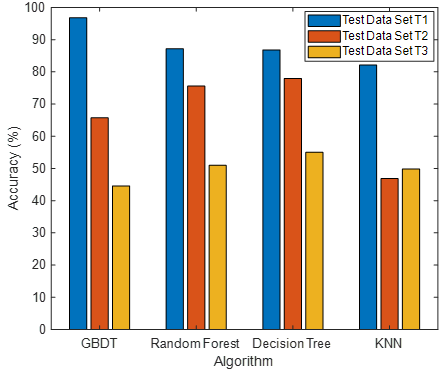}
  \caption{Classification accuracy of each algorithm for the three test data sets}
  \label{fig:Accuracy}
\end{figure}

As shown in Fig.~\ref{fig:Dataset}, the machine learning model was trained using the data set T0, and the trained model was tested using the data sets T1, T2, and T3.
Fig.~\ref{fig:Accuracy} shows the classification accuracy of each machine learning algorithm for the test data set T1, T2, and T3.
When the trained model was tested using the data set T1, the classification accuracy was 82\%–96\%, which was significantly high.
This was because the training and test data sets, T0 and T1, respectively, consisted of data collected at the same locations (that is, locations A, B, and C).
However, when the trained model was tested using the test data sets T2 and T3, which were collected at locations different from the training data locations (that is, locations D and E), classification accuracies of 46\%–77\%, and 44\%–55\% were obtained, respectively. The decision tree algorithm performed better than other algorithms in this case.

\begin{figure}
  \centering
  \includegraphics[width=0.8\linewidth]{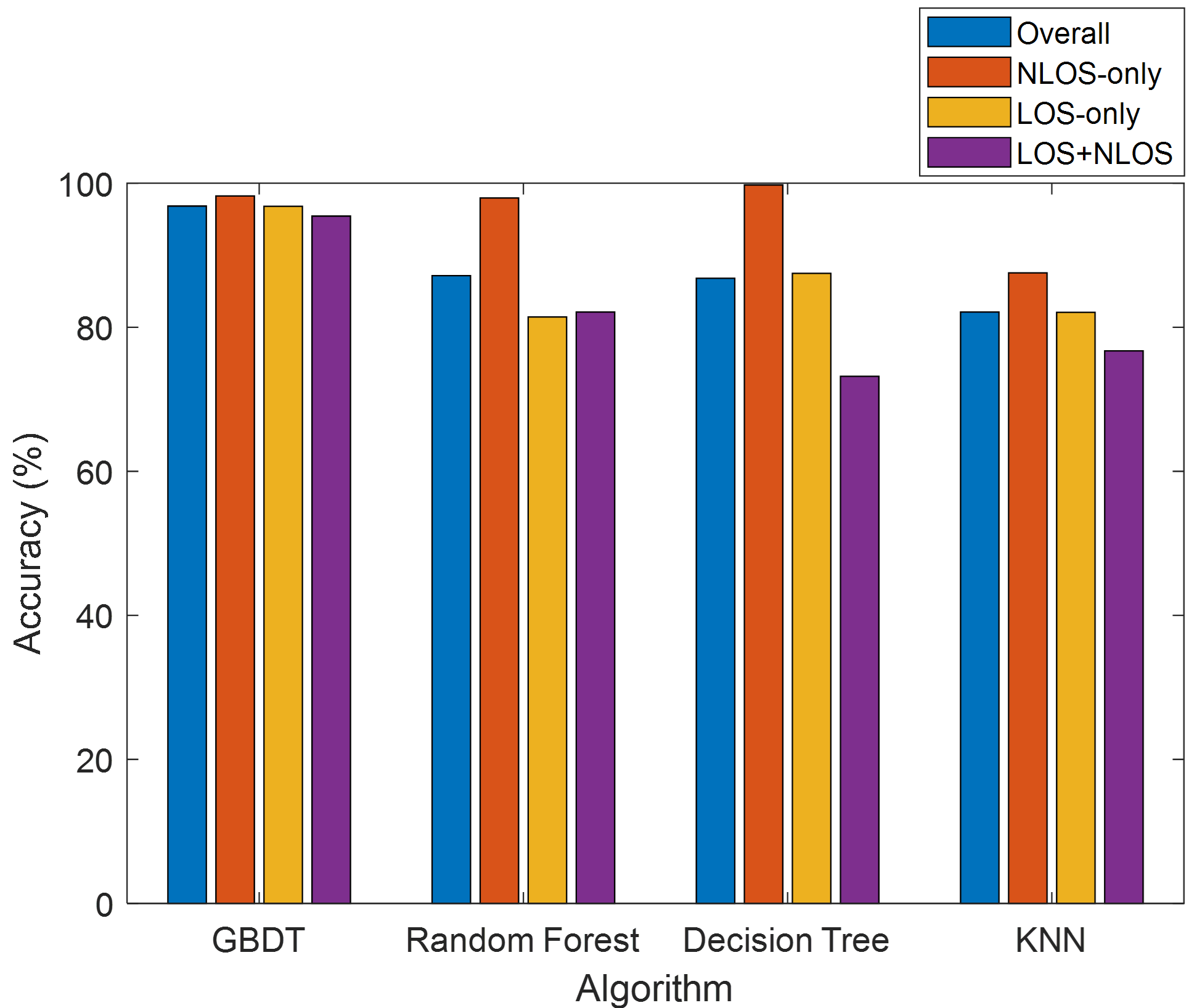}
  \caption{Classification accuracy of each algorithm according to the signal reception conditions in the test data set T1}
  \label{fig:AccT1}
\end{figure}

\begin{figure}
  \centering
  \includegraphics[width=0.8\linewidth]{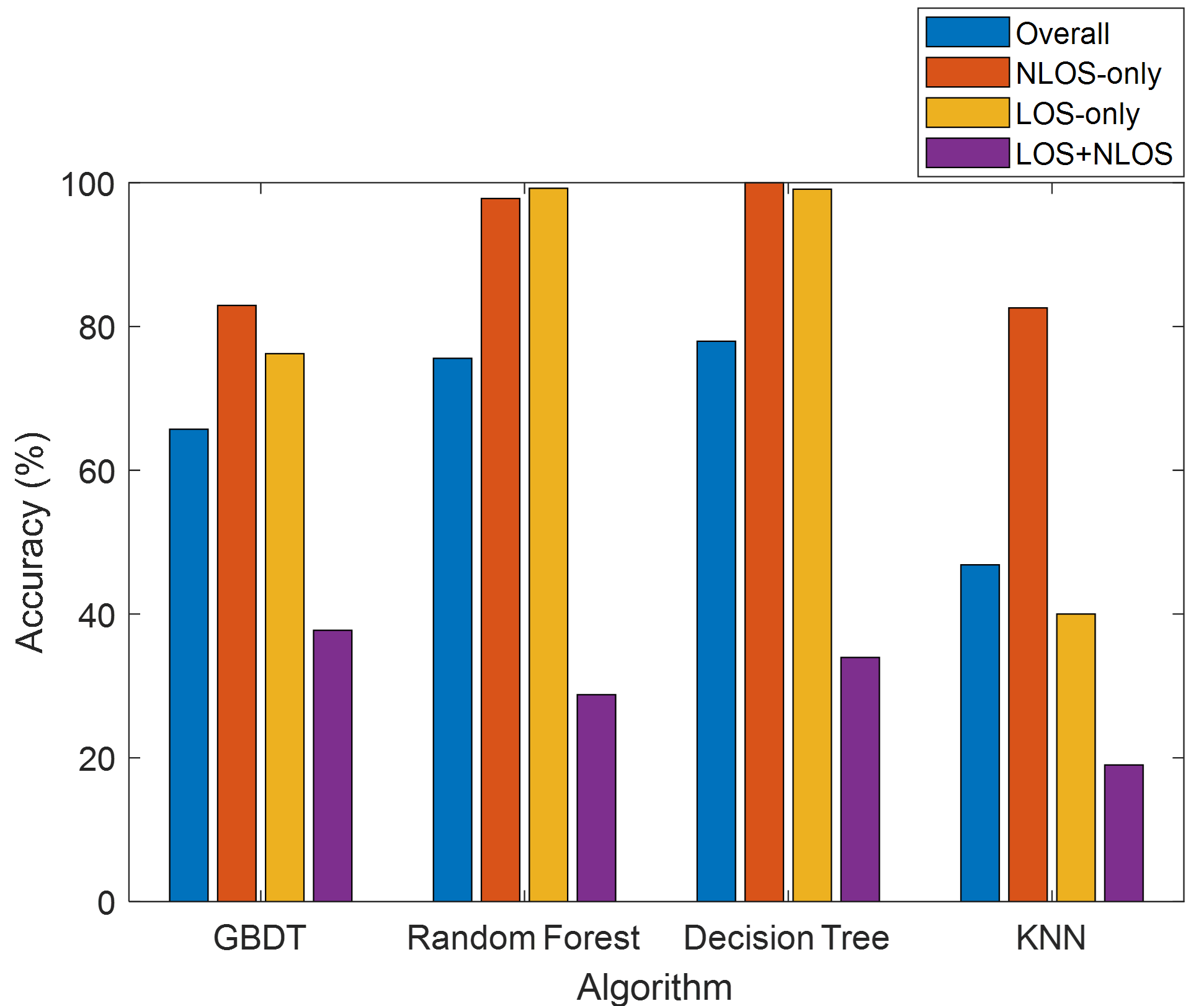}
  \caption{Classification accuracy of each algorithm according to the signal reception conditions in the test data set T2}
  \label{fig:AccT2}
\end{figure}

\begin{figure}
  \centering
  \includegraphics[width=0.8\linewidth]{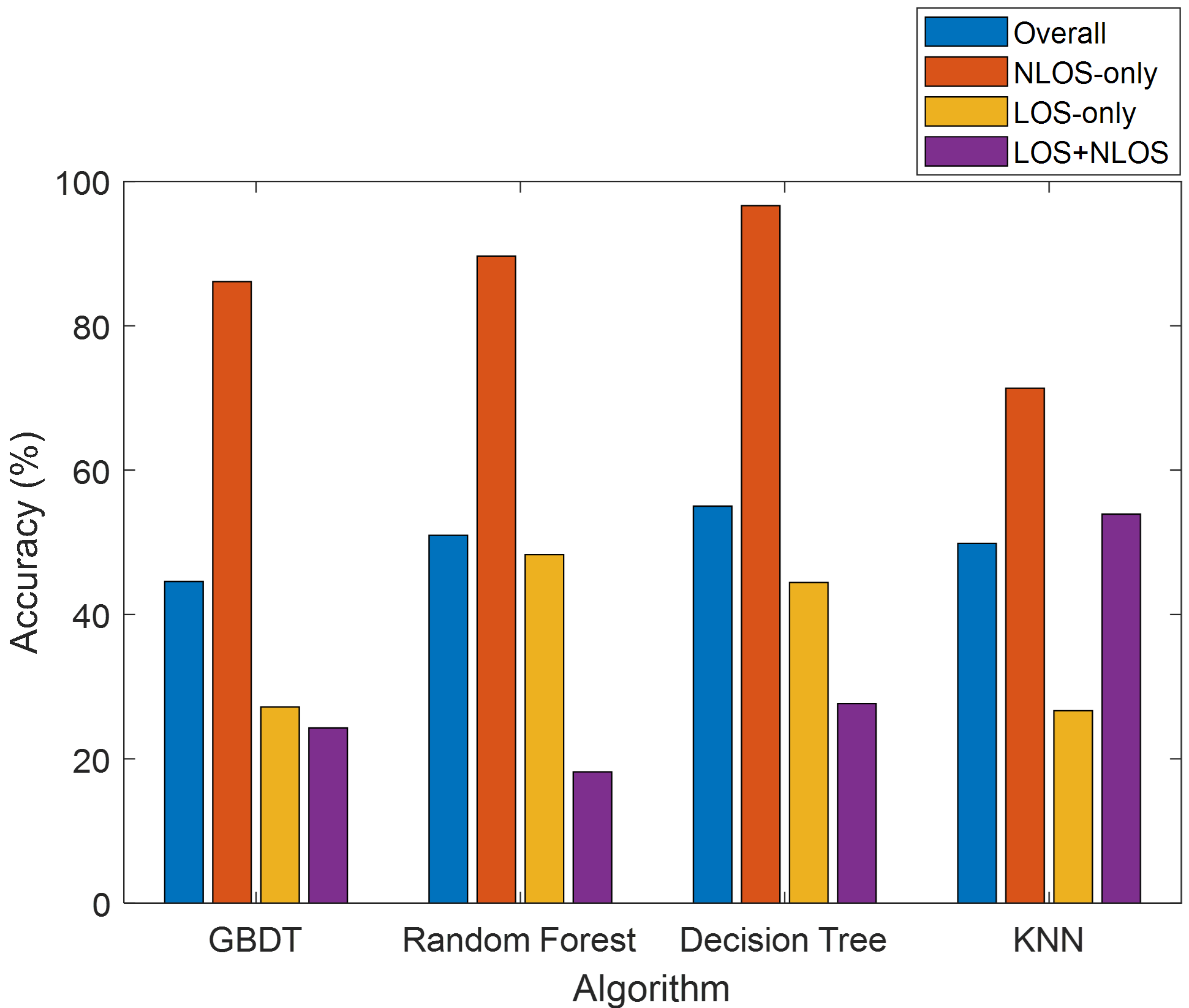}
  \caption{Classification accuracy of each algorithm according to the signal reception conditions in the test data set T3}
  \label{fig:AccT3}
\end{figure}

Figs.~\ref{fig:AccT1}--\ref{fig:AccT3} show the classification accuracy of each machine learning algorithm according to the signal reception conditions.
For the data set T1, as shown in Fig.~\ref{fig:AccT1}, an accuracy of 70\% or greater was achieved.
However, for the data sets T2 and T3, as shown in Figs.~\ref{fig:AccT2} and  \ref{fig:AccT3}, respectively, the classification accuracy of the LOS+NLOS signals was only 18\%–53\%, which was significantly lower than that of the other cases.
This result implied that the training data set should include more LOS+NLOS data samples to achieve a better training result. This was understandable because the LOS+NLOS environment had greater diversity than the LOS-only or NLOS-only environments, and thus required more training.

\section{Conclusion}

In this study, we proposed a machine learning-based GPS multipath detection method that uses dual antennas.
We used five features for machine learning, comprising four features that could be obtained from a single antenna, and one feature that could be obtained from dual antennas.
The performances of four machine learning algorithms (that is, GBDT, random forest, decision tree, and KNN) using the proposed features were evaluated.
The classification accuracy for LOS+NLOS signals was found to be low. However, it is expected to improve if more LOS+NLOS data samples are used for the training, and this can be investigated in a future work.

\bibliographystyle{IEEEtran}
\bibliography{mybibfile, IUS_publications}

\end{document}